\documentclass[12pt, draftclsnofoot, onecolumn]{IEEEtranTCOM}
\renewcommand{\baselinestretch}{1.5}

\usepackage{indentfirst}
\usepackage{multirow}
\usepackage{graphicx}
\usepackage{color}
\usepackage{xcolor}
\usepackage{colortbl}
\usepackage{listings}
\usepackage{enumerate}
\usepackage{subfigure}
\usepackage{amsmath,amsthm,amscd,amssymb,latexsym,upref,stmaryrd}
\usepackage[noend]{algpseudocode}
\usepackage{algorithmicx,algorithm}
\usepackage{amsfonts,hyperref,epsfig}
\usepackage{CJK}
\usepackage{float}
\usepackage{cite}
\usepackage{subfigure}
\usepackage{lipsum}
\usepackage{multicol}
\usepackage{setspace}
\usepackage{bm}
\usepackage{stfloats}
\usepackage{geometry}
\usepackage{mathrsfs}
\begin{document}
\title{Integrated Communication and Localization \\ in mmWave Systems}

\author{Jie~Yang, Jing~Xu, Xiao~Li, Shi~Jin, and Bo~Gao
	\thanks{Jie~Yang, Jing~Xu, Xiao~Li, and Shi~Jin are with the National Mobile Communications Research Laboratory, Southeast University, Nanjing, China (e-mail: \{yangjie;shadowaccountxj@foxmail.com;li\_xiao;jinshi\}@seu.edu.cn). Bo~Gao is with the ZTE Corporation and the State Key Laboratory of Mobile Network and Mobile Multimedia Technology, Shenzhen, China (e-mail: gao.bo1@zte.com.cn).}
}
	
\maketitle	

\begin{abstract}
As the fifth-generation (5G) mobile communication system is being commercialized, extensive studies on the evolution of 5G and sixth-generation mobile communication systems have been conducted.
Future mobile communication systems are evidently evolving towards a more intelligent and software-reconfigurable functionality paradigm that can provide ubiquitous communication and also sense, control, and optimize wireless environments.	
Thus, integrating communication and localization by utilizing the highly directional transmission characteristics of millimeter-wave (mmWave) is a promising route.
This approach not only expands the localization capabilities of a communication system but also provides new concepts and opportunities to enhance communication.	
In this paper, we explain the integrated communication and localization in mmWave systems, in which these processes share the same set of hardware architecture and algorithm. 
We also perform an overview of the key enabling technologies and the basic knowledge on localization.
Then, we provide two promising directions for studies on localization with extremely large antenna array and model-based neural networks.
We also discuss a comprehensive guidance for  location-assisted mmWave communications in terms of channel estimation, channel state information feedback, beam tracking, synchronization, interference control, resource allocation, and user selection.  
Finally, this paper outlines the future trends on the mutual assistance and enhancement of communication and localization in integrated systems.	
\end{abstract}

\begin{IEEEkeywords}
millimeter wave, integrated communication and localization, location-assisted communication, extremely large antenna array, reconfigurable intelligent surface, artificial intelligence, neural networks
\end{IEEEkeywords}

\section{Introduction} \label{sec:introduction}

Millimeter-wave (mmWave) band communications are expected to play a pivotal role in the fifth-generation (5G) and upcoming sixth-generation (6G) mobile communication systems \cite{6824752,6736746,8732419}.
The mmWave band offers an extremely wide bandwidth and can boost peak data rates. 
However, mmWave signals have inherently large path-loss and are sensitive to blockage.
Hence, large antenna arrays and highly directional transmission should be combined to compensate for the severe penetration path-loss \cite{6834753,heath,7959169}.
Emerging techniques, such as extremely-large antenna array (ELAA) \cite{elaa3,elaa1,elaa2}, ultra-dense networks (UDN) \cite{udn}, reconfigurable intelligent surface (RIS) \cite{ris2,ris3,ris1}, and artificial intelligence (AI) \cite{DL,TQ,xu,model-driven},  further enhance the coverage and capability of the mmWave communication system.
Hence,  providing ubiquitous connectivity with ultra-high throughput, ultra-low latency, and ultra-high reliability is promising \cite{6G1,6G}.

The highly directional transmission makes localization a greatly desirable feature of mmWave communication systems \cite{loc1}.
Although the location is available from the second-generation to 5G mobile communication systems, the accuracy is limited to the range of several hundreds to several meters \cite{8226757}.
The aforementioned key technology enablers will empower the mmWave communication systems with very fine range, Doppler, and angular resolutions to realize a centimeter-level localization.
Therefore, localization will be built-in and will reuse the infrastructure and resources that are used for wireless communications.
Thus, localization is easy and cost-effective to deploy and beneficial to support flexible and seamless connectivity.
The highly accurate location information can be used to provide location-based services, such as navigation, mapping, social networking, augmented reality, and intelligent transportation systems. 
Additionally, location-assisted (or location-aware) communications can be realized by the obtained location information to improve the communication capacity and network efficiency \cite{6924849}.

Although considerable advantages in integrating communication and localization in mmWave systems have been predicted,
studies in this field have just started \cite{ILC,heathlc}.
Localization and communication are still studied separately.
Integrating localization and communication in mmWave systems 
will disrupt the traditional algorithm and protocol design and revolutionize the way of communication.
Many challenges need to be solved urgently, including the exchange of information, sharing resources, and performance tradeoffs between communication and localization. 
Xiao and Zeng \cite{ILC} presented a holistic survey on the basis of wireless localization and how wireless localization and communication interplay with each other in various network layers.
Ali et al. \cite{heathlc} motivated the
use of infrastructure-mounted sensors (which will
be part of future smart cities) to aid in establishing
and maintaining mmWave vehicular communication links. 
In this paper, we present a progressive vision on integrating communication and localization by sharing the same mmWave hardware and software architectures.
We discuss in detail the advancement in the architecture and algorithm design of mmWave communication systems for localization
and how the location assists the communication links in different ways.  
We also summarize some promising future research directions to help materialize the integrated communication and localization in mmWave systems over the next few years.

The rest of this paper is organized as follows. In Section \ref{s2}, we explain the concept of the integrated communication and localization. We also review the background of signal and channel models of mmWave communications, which are the basis of communication and provide the necessary parameters and information for localization. We discuss the key enabling technologies that enhance the communication and localization in mmWave systems.
In Section \ref{s3}, we focus on the localization techniques in an mmWave communication system by overviewing the basic knowledge and proposing two promising research directions, i.e., localization with extremely large antenna array and model-based neural networks, on the basis of the new hardware architecture and algorithm system, respectively.
We discuss the location-assisted mmWave communications in Section \ref{s4} in terms of channel estimation, CSI feedback, beam tracking, synchronization, interference control, resource allocation, and user selection. 
Finally, we outline the future trends in Section \ref{s5} and conclude the paper in Section \ref{s6}.

\section{Integrated Communication and Localization}\label{s2}
The mmWave, massive multiple-input multiple-output (MIMO), and UDN techniques have driven the development of low-complexity transceiver architectures and new signal processing techniques \cite{6834753,s12,heath,s11,7959169,8387211,yx,yj1,yj2}.
The evolution of this technology has inspired numerous potential applications in mmWave communication systems, such as automatic vehicle and augmented reality. 
Highly-accurate localization can also be achieved in mmWave MIMO communication systems by leveraging distinguishable multipath components provided by the mmWave channel, without installing additional dedicated infrastructure \cite{loc1,mmwave1,mmwave2,loc2,slam2}.
The mmWave bands offer larger bandwidths than the presently used sub-6 GHz bands.
Hence, higher resolution of the time of arrival (TOA) and frequency of arrival (FOA) can be achieved. 
In addition, large antenna arrays and highly directional transmission enhance the resolution of the angle of arrival (AOA) and angle of departure (AOD).
The sparsity of the mmWave channel in beamspace can be utilized to reduce the complexity of signal processing and cost of hardware, which also simplifies the elimination of non-line-of-sight (NLOS) path interference for localization.
Moreover, by leveraging distinguishable multipath components and geometric relationships between the NLOS paths and scatterers, mmWave MIMO communication systems can turn multipath channels ``from foe to friend''.
Therefore, the localization in mmWave communication systems is expected to achieve centimeter-level accuracy.

However, to fully realize high-throughput wireless communications, many challenges in the mmWave bands need to be addressed.
For instance, large path-loss and high blockage probability are still key factors restricting the development of mmWave communications, especially in high-mobility scenarios.
Therefore, frequent beam training and effective beam tracking are indispensable to maintain directional communications, and these processes complicate the link establishment and introduce a large overhead.
One possible solution to support high mobility is integrating communication and localization in mmWave communication systems.
Therefore, the location can be used more flexibly to assist the beam training and prediction processes.

\begin{figure}
	\centering
	\includegraphics[scale=0.65]{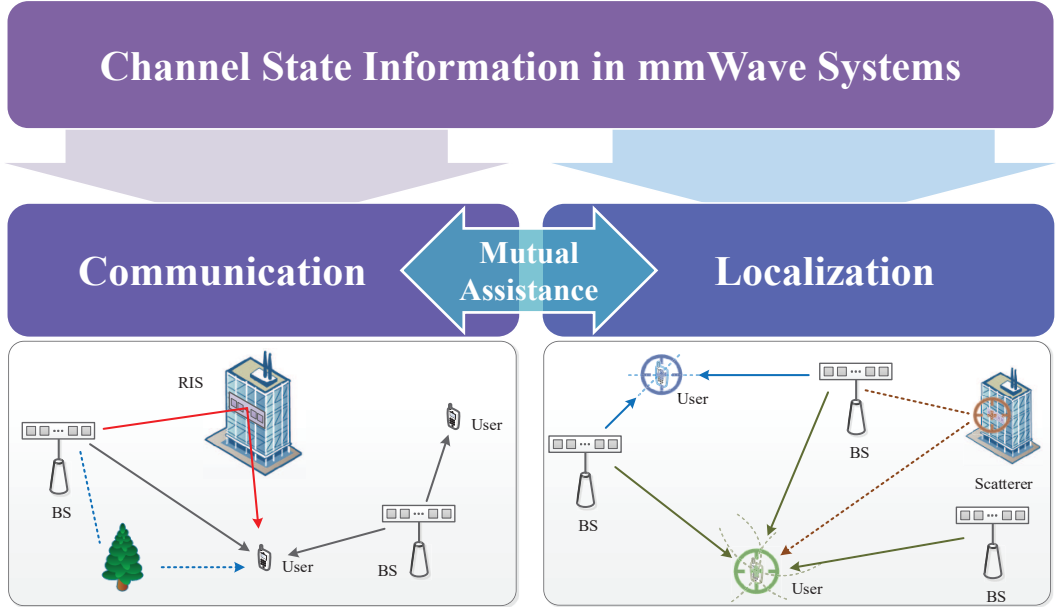}
	\caption{Integrated communication and localization in millimeter-wave systems} \label{fig:ILC}
\end{figure}

\textbf{The integrated communication and localization technique} involves a high degree of integration of advanced technologies in communication and localization at the level of hardware architecture and algorithm system by sharing the infrastructure and time-frequency-space resources of wireless communications.
Coordinating communication and localization can be empowered by the information interaction capability of the high-rate and low-delay mmWave communication systems.
The joint design of communication and localization breaks the traditional pattern of separate operations and achieves high-throughput communications and highly-accurate localizations in one system.
Hence, the channel state information (CSI) obtained by channel estimation is not only the basis of communication but also captures the side information of displacements and movements of the transmitter, receiver, and surrounding scatterers.
As shown in Fig. \ref{fig:ILC}, the integrated communication and localization in the mmWave systems is based on the CSI or CSI-related parameters.
Then, the mutual assistance and enhancement of communication and localization can be realized iteratively, in which more reliable communication provides more accurate measurements required by localization. 
In addition, more precise estimation of location reduces communication overhead.

\subsection{Signal and Channel Models}

In a MIMO wireless communication system with $N_T$ transmit antennas and $N_R$ receive antennas, 
the received signal at time $t$ can be expressed as follows:
\begin{equation}\label{receiver}
\mathbf{y}(t)=\mathbf{H}(t)\ast\mathbf{x}(t)+\mathbf{n}(t),
\end{equation}
where $\mathbf{H}(t) \in \mathbb{C}^{N_R \times N_T}$ is the channel impulse response matrix, $\mathbf{x}(t)\in \mathbb{C}^{N_T \times 1}$ is the transmitted signal vector, $\mathbf{n}(t)\in \mathbb{C}^{N_T \times 1}$ is the noise vector, and $\ast$ denotes the convolution operator. 
In wireless communications, the electromagnetic wave, which carries the communication information, interacts with the propagation environment in accordance to Maxwell's equations.
For example, the received signal strength (RSS) is dependent on the path-loss model, which roughly reflects the distance between the receiver and the transmitter.
Given that the RSS is easily obtained, this parameter has been widely used in localization methods based on either trilateration or fingerprinting.
However, 
the unstable fluctuating nature of RSS deteriorates the accuracy of localization.
In particular, the RSS is affected by small-scale fading and body shadowing.  
Hence, we focus the fine-grained information provided by CSI,
in which the CSI amplitude and phase are impacted by the displacements and movements of the transmitter, receiver, and surrounding scatterers.
Therefore, the CSI can be used not only for communication but also localization via mathematical modeling or machine learning algorithms.

The multi-path time-varying channel impulse response for an mmWave MIMO system is given as follows \cite{MIMO,LTV}:  
\begin{equation}\label{cir}
\mathbf{H}(t)= \sum_{l=0}^{L}\alpha_{l}\mathbf{a}_{\text{R}}(\theta_{l}^{\text{az}},\theta_{l}^{\text{el}})\mathbf{a}_{\text{T}}^T(\phi_{l}^{\text{az}},\phi_{l}^{\text{el}})\delta(t-\tau_{l})e^{j2\pi \nu_{l} t},
\end{equation}
where $L$ is the total number of discrete propagation paths;
$\alpha_{l}$, $(\theta_{l}^{\text{az}},\theta_{l}^{\text{el}})$, $(\phi_{l}^{\text{az}},\phi_{l}^{\text{el}})$, $\tau_{l}$, and $\nu_{l}$ are the complex attenuation factor, AOAs, AODs, TOA, and Doppler frequency associated with the $l$-th path, respectively; and
$\mathbf{a}_{\text{R}}(\cdot) \in \mathbb{C}^{N_R \times 1}$ and $\mathbf{a}_{\text{T}}(\cdot) \in \mathbb{C}^{N_T \times 1}$ are the steering vectors of the receive and transmit antenna arrays \cite{sv}, respectively.
Time-frequency-spatial domain CSIs are provided by the mmWave communication systems, with orthogonal frequency-division multiplexing and orthogonal time-frequency-space modulation.
According to \eqref{cir}, the multi-path information needed for localization, e.g., AOAs, AODs, TOAs, and Doppler frequencies, can be effectively obtained by advanced parameterized channel estimation algorithms \cite{7491265,7827161,8697125}.

\subsection{Enabling Technologies}
Various emerging technologies, such as ELAA, RIS, and AI are extensively discussed as promising technologies to boost mmWave communication capabilities with high-throughput, massive-connectivity, high-reliability, and low-latency.
At the same time, the advanced enabling technologies are enhancing the localization capability of mmWave communication systems.

\textbf{ELAA} techniques drive the antenna deployment towards larger apertures and greater numbers than those of commercial cellular systems, and this phenomenon can boost spatial diversity and enhance communication coverage.
Moreover, as the antenna dimension continues to increase, the user-equipment and significant scatterers are likely to be
located in the near-field of the array. Consequently, the spherical wave phenomenon emerges.
The inherent phase shifts in the space caused by spherical waves provide new opportunities for localization in ELAA systems \cite{s1,s6}.

\textbf{RIS} consists of tunable unit cells and has recently drawn significant attention because of its superior capability in manipulating electromagnetic waves. In particular, RIS-assisted wireless communications have the great potential to enhance significantly the performance and coverage in a cost-effective and energy-efficient manner by the proper programming of the reflection coefficients of the unit cells. 
Therefore, RIS is promising in extending the wireless communication range, facilitating NLOS communications, and providing low-cost cooperative localization opportunities \cite{locriss}.

\textbf{AI} has received considerable attention because
of its promising learning ability in solving complex problems, which can alleviate modeling issues.
Deep learning has been widely recognized as the current representative general advances of AI and has shown performance advantages in many aspects.
The state-of-the-art computer vision techniques can be embedded in the integrated mmWave communication and localization systems by considering CSI matrices as images. 
Learning methods based on hybrid data and models have also emerged to reduce the training data and enhance the robustness of the algorithm \cite{xu,model-driven}.

The mentioned key technology enablers present exciting new opportunities for the integration of communication and localization in mmWave systems.
This phenomenon bring challenges to the traditional design principles.
The introduction of ELAA and RIS will invalidate the traditional channel model.
New channel models are required to be dynamic and consistent across frequencies, space, and materials.
Moreover, the use of thousands or more active antenna elements will generate a prohibitive cost in terms of hardware implementation, energy consumption, and complexity of signal processing. Subarray and beamspace hardware architectures and the corresponding signal processing techniques need further enhancement.
Moreover, real-time energy-efficient AI/ML techniques should be developed to achieve high-throughput communication and highly-accurate localization with limited data and in adaptive environments.
To achieve high-throughput and low-latency communication together with centimeter-level localization accuracy, 
the mutual assistance and enhancement between communication and localization in the integrated mmWave system becomes essential.

\section{Localization in the mmWave Communication Systems}\label{s3}

Localization in a mmWave communication system is aimed to estimate the location, velocity, and orientation of the \textit{user-equipment} or \textit{agent nodes} together with possible scatterers on the basis of a set of wireless reference signals transmitted or received by the \textit{base-stations} or \textit{anchor nodes}.
Deployment of localization by reusing the infrastructure that is used for mmWave communications is convenient and cost-effective.
This process recycles real-time CSI that is already processed at the receiver in communication systems, as shown in Fig. \ref{fig:ILC}.
Moreover, contrary to solutions based on sensor, video, and WiFi, localization in mmWave communication systems is not intrusive and insensitive to lighting conditions. In addition, this process has unprecedented time-frequency-spatial resolution.
The high-throughput offered by mmWave communication links can be leveraged to quickly and reliably share map and location information among different devices.
Localization techniques can be categorized from various
perspectives, as summarized in \cite{sumloc,ILC}, and popular classifications are listed as follows:

(1) Direct and indirect localization. 
\textit{Direct localization} converts directly the received waveform to estimate a location, avoiding the error propagation but involving a highly complex process.  
By contrast, \textit{indirect localization}, which is more popular, applies the principle in
which the channel parameters, such as RSS, AOA, TOA, time difference of arrival (TDOA), FOA, and frequency difference of arrival (FDoA), are first extracted from the received waveform and grouped as a function of the location parameters.
Then different estimators are used to
determine the user positions.
However, the intermediate error carried by the two-step method deteriorates the accuracy of localization.

(2) Uplink and downlink localization.
\textit{Uplink localization} is performed in the base-station or at the central unit.
This process is also named network-based and cloud-based localization.
All time-consuming and complex computing operations are performed at the base-station or the central unit, which is appealing for resource-limited agent devices.
The location of all agent nodes is shared by anchor networks, which facilitate location-aware communications. 
However, the privacy and security of the location are critical issues.
\textit{Downlink localization} also named device-based and edge-based localization, is executed on the agent nodes, and this process has inherent mechanism to protect the privacy of users.
However, the agent nodes have high hardware requirements for accurate localization.

(3) Model-based and data-based localization.
\textit{Model-based localization} exploits the geometric properties of the angle and range measurements to locate the agent nodes, e.g., multilateration and multiangulation.
A serious flaw of model-based localization lies in the difficulty in accurately modeling complex multi-path environments.
\textit{Data-based localization} utilizes database matching methods or standard neural network architectures to ease modeling issues. This process requires offline training step to extract unique geotagged signatures from the collected data in the area of interest and online localization step to match online measurements against the pre-recorded signatures in the database.
The main limitations are the high time-and-resource consumption and the lack of theoretical guidance of the neural network design.

In the following subsections, we provide two promising research directions of localization in the mmWave communication systems, namely, the new hardware architecture and the new algorithm system.

\subsection{Localization with Extremely Large Antenna Array}

The radiation field of an antenna array is divided into the near-field and far-field regions via the Rayleigh distance \cite{nearfiled,sv}, which is given as 
$
R = {2D^2}/{\lambda},
$
where $D$ is the maximum dimension of the antenna array, and $\lambda$ is the wavelength. 
When the distance between the user (or scatterer) and the base-station is smaller than the Rayleigh distance, the user (or scatterer) is located in the near-field region, where the spherical wave-front over the antenna array is observed.
For example, a uniform linear array (ULA) of 1 m that operates at $30$ GHz corresponds to a Rayleigh distance of approximately $200$ m and nullifies the uniform plane wave-front model usually assumed in previous studies on wireless communications. 
As the antenna dimension continues to increase, the range of the radiative near-field of the antenna array expands, and the user and significant scatterers are likely to be located in the near-field of the array.

\begin{figure}[tbh]
	\centering
	\includegraphics[scale=0.6]{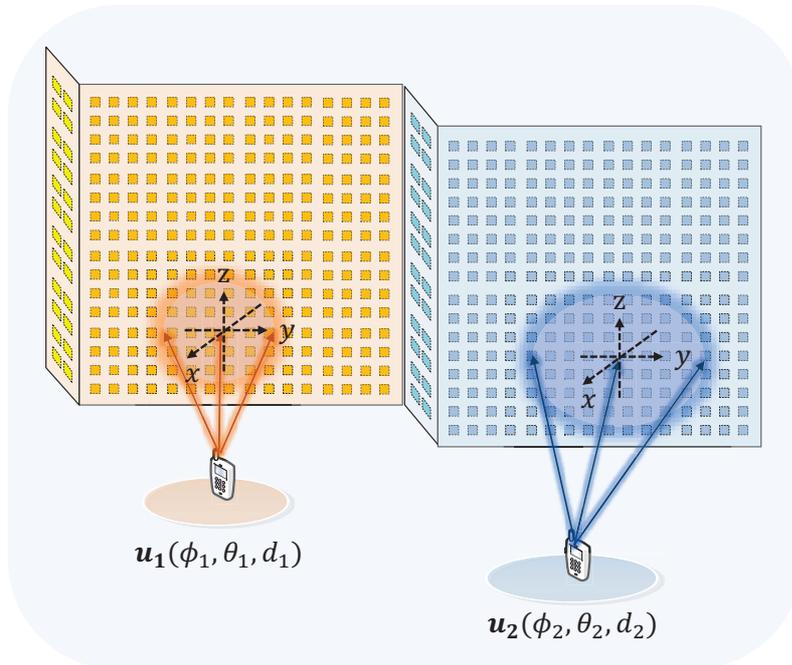}
	\caption{Localization with extremely large antenna arrays} \label{fig:elaa}
\end{figure}

According to \cite{s3} and its references,
the standard response model for the large ULA with spherical-wave is as follows:
\begin{equation}\label{spherical}
\mathbf{a}_s(\phi,\theta,d)= \left[
\begin{matrix} 
\frac{d}{\tilde{d}_1}e^{-j\frac{2\pi}{\lambda}(\tilde{d}_1-d)}\\
\frac{d}{\tilde{d}_2}e^{-j\frac{2\pi}{\lambda}(\tilde{d}_2-d)}\\
\cdots \\
\frac{d}{\tilde{d}_m}e^{-j\frac{2\pi}{\lambda}(\tilde{d}_m-d)}
\end{matrix} \right],
\end{equation}
where $\phi$ and $\theta$ denote the azimuth and elevation angles of the source with respect to the reference point, respectively;
$d$ is the distance from the source to the reference point;
and $\tilde{d}_m$ is the distance from the source to the $m$-th antennas.
With fixed space between antennas, $\tilde{d}_m$ is determined by index $m$ and parameters $(\phi,\theta,d)$.
Equation \eqref{spherical} is compatible with conventional plane wave models. 
The underlying parametric model \eqref{spherical} allows the characterization of a path with a new parameter, i.e., the distance between the source and the reference point, additional to the conventional parameters characterizing a path under the plane-wave assumption.
Therefore, the near-field effects facilitate the exploitation of the wave front curvature to jointly estimate the range and direction of the source.
This process can improve the accuracy of location and possibly remove the need for explicit synchronization between reference anchors (Fig. \ref{fig:elaa}).

Several studies have started to investigate the localization potential with advanced large antenna arrays to realize joint communication, control, and localization in mmWave communication systems \cite{s4,s5,s3,s6,s7}.
An algorithm based on maximizing the space-alternating generalized expectation to estimate the locations of scatterers in the last-hops of the propagation paths is proposed in \cite{s4}, in which a large-scale ULA is used in a receiver to measure the channel. 
The advantages of near-field in localization and synchronization over far-field is shown in \cite{s5} from a Fisher information
perspective.
\cite{s3} claims the limitations of the standard array response model \eqref{spherical}, which sometimes significantly differs from the model based on electromagnetic theory and does not consider the characteristics of the near-field source, such as the type and orientation of the transmitting antennas.
These characteristics may have a profound impact on the signals received by the array.
More accurate models are required when attempting to perform high-resolution localization of closely spaced signals.
In addition to ULA and UPA, the potential for positioning with a large intelligent surface or RIS \cite{s6} and lens antenna array \cite{s7} have been exploited. \cite{s6} derived the closed-form Fisher information matrix and Cram\'{e}r-Rao lower bounds to position a terminal with and without the unknown phase $\phi$ presented in the analog circuits of the RIS. \cite{s7} investigated the possibility of directly positioning with a single large lens antenna array by retrieving information from the wave front curvature.

The use of thousands or more active antenna elements in the ELAA systems will generate prohibitive cost in terms of hardware implementation, energy consumption, and complexity of signal processing \cite{s10}.
One effective solution to significantly reduce the complexity of the system and cost of implementation caused by the large number of active antennas and users is to partition the antenna array into a few disjoint subarrays \cite{elaa1,s8,s9}.
Another method is by using the energy-focusing property of an {ex}tremely large {lens} antenna array denoted as ``ExLens'', which can fully utilize the aperture offered by large antenna arrays. 
The complexity of signal processing and cost of the radio frequency (RF) chain could be significantly reduced without notable degradation of performance in mmWave and massive MIMO systems by utilizing lens antenna arrays \cite{s12,s11}.
Electromagnetic lenses can provide variable phase shifting for electromagnetic rays at different points on the lens aperture to achieve angle-dependent energy focusing property.
Therefore, lens antenna arrays can transform the signal from the antenna space to the beamspace (the latter has lower dimensions) to reduce the RF chains significantly.
Recent studies have confirmed that the beamspace is one of the promising enablers for communication and localization in mmWave communication systems \cite{6G1}.

\subsection{Localization with Model-based Neural Networks}

\begin{figure}[tbh]
	\centering
	\includegraphics[scale=0.5]{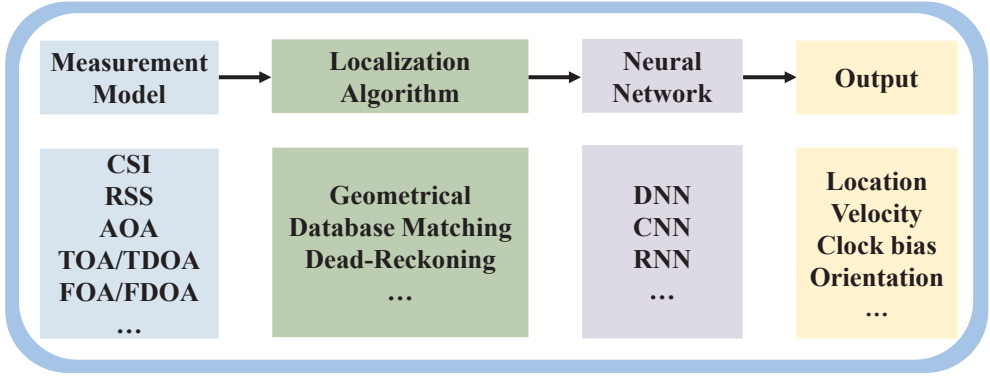}
	\caption{Localization with model-based neural networks} \label{fig:MNN}
\end{figure}

To overcome the disadvantages of localization methods bassed on pure data or model, we propose a localization method based on hybrid data and model, i.e., the localization with model-based neural networks.
With the proposed technique, neural network topology with theoretical localization foundations can be designed, and the network structure can be explained and predicted \cite{xu,model-driven}.
At present, the localization by combining neural networks with geometric models is rarely reported.
The model-based neural network approach obviously retains the advantages of the model-based approach (determinacy and theoretical soundness) and powerful learning ability of the data-based approach.
It also overcomes the difficulties in accurate modeling and avoids the large requirement for time and computing resources.
Localization with model-based neural networks comprises three parts,  \textbf{measurement model}, \textbf{localization algorithm}, and \textbf{neural network}, as shown in Fig. \ref{fig:MNN}. 

\textbf{Measurement models} are constructed based on the domain knowledge in localization developed over several decades of intense research, but only need to provide a very rough and broad definition of the solution, thereby reducing the pressure for accurate modeling.
In particular, the localization model can be established by utilizing the CSI matrix \cite{csi1,csi2}, RSS \cite{rss}, AOA \cite{aoa}, TOA, TDOA, FOA, and FDOA measurements \cite{toa,fdoa}.
Channel parameters related to RSS, AOA, TOA, TDOA, FOA, and FDOA have the underlying geometric relationship with location parameters. For example,
for the LOS path, the relationship is as follows:
\begin{equation} \label{1}
\tau_{0}=\dfrac{||\mathbf{u}-\mathbf{b}||}{v_c}+\omega,
\end{equation}
where $\tau_{0}$ is the TOA of the LOS path; $\mathbf{u}$ is the location of the user; $\mathbf{b}$ is the location of the base-station; $v_c$ is the signal propagation speed; and $\omega$ is the unknown clock bias between the base-station and user.	
For the single-bounce NLOS path, the relationship is as follows:
\begin{equation} \label{2}
\tau_{l} = \dfrac{||\mathbf{u}-\mathbf{s}_{l}|| + ||\mathbf{s}_{l} - \mathbf{b}||}{v_c}+\omega,
\end{equation}
where $\tau_{l}$ is the TOA of the $l$-th NLOS path, and $\mathbf{s}_{l}$ is the location of the $l$-th scatterer.
The selection of measurements is a tradeoff between performance
and cost. 
RSS can be conveniently collected without extra hardware.
However, the localization accuracy is low, because the parameters of the path-loss model are unknown and fluctuate in complex environments.
Some measurements, such as AOA, TOA, TDOA, FOA, and FDOA, can be used to achieve high localization accuracy but require extra hardware or modifications of nodes.
The AOA requires a large antenna array and advanced phase detection for accurate angular-measurement.  
The TOA needs precise timing among users and base-stations.
It is also common to use hybrid measurements for better localization performance.
Moreover, data from sensors, such as gyroscope, magnetometer, accelerator, and map, can be fused with wireless signals to enhance the resolution of localization.
All measurements suffer from deterministic and stochastic measurement errors. 
The deterministic errors can be calibrated by offline or online methods.
The stochastic errors should be modeled as stochastic processes.

\textbf{Localization algorithms} are designed to solve the localization problems based on the aforementioned model and domain knowledge.
The localization tasks are not trivial given that the relations are nonlinear and nonconvex functions of $\mathbf{u}$, $\dot{\mathbf{u}}$, $\mathbf{s}_{l}$, and $\dot{\mathbf{s}}_{l}$.
Main localization algorithms are geometrical algorithms, database matching, and dead-reckoning, as summarized in \cite{non-linear,sumloc,ILC,locsum2}. 
The geometrical localization algorithms include multilateration \cite{toa}, multiangulation \cite{aoa}, and multiangulateration \cite{aoatoa}.
These algorithms mainly involve nonlinear and linear location estimators.
The nonlinear estimators directly solve the
problems by minimizing a cost function, such as nonlinear least squares, weighted nonlinear least squares, and maximum likelihood estimators.
These nonlinear estimators usually have satisfactory localization
accuracy but are time-consuming if the grid or random search is applied.
By contrast, linear estimators, such as linear least squares and weighted linear least squares estimators, can approximate the nonlinear equations into a set of linear equations.
Thus, these estimators can find efficient solutions quickly but with degraded localization accuracy compared with nonlinear estimators. 
Database matching algorithms use the measurements as fingerprints to find the closest match, and these algorithms are suitable for complex scenarios that are difficult to parameterize \cite{fp,fp2}.
The geometrical and database localization algorithms can be integrated to reduce the consumption of resources and computational load, thereby achieving a robust performance \cite{crowdsourcing,mfp}.
Recent developments in deep learning have resulted in its great potential for the integrated model-based and data-based localization methods \cite{xu,model-driven}. 

\textbf{Neural networks} are generally constructed by unfolding an iterative algorithm to deep neural networks and  replacing the linear approximate operations in the model by classic neural networks.
For a preprint \cite{yang}, localization by neural network-assisted weighted linear least squares is established by introducing neural networks into the developed weighted linear least squares model to learn higher-order error components.
Thus, the performance of the estimator is improved, especially in a highly noisy environment.
Introducing learnable parameters is also a feasible mechanism.
For example, the parameters in the RSS model are uncertain and closely coupled with the environment.
Therefore, neural networks can be embedded to learn particular parameters. 
The environment adaptability of the algorithms can be improved by adding a small amount of training data.
Hence, the depth of the neural networks is determined by the estimating convergence rate of the algorithm family. 
The parameter space of the deep network is determined by the parameter constraints.
The topology of the deep network is determined by the algorithms, 
and the deep network can be trained by back-propagation.
In addition, unsupervised learning, reinforcement learning, and federated learning can be organically combined with the localization models.

\section{Location-assisted mmWave Communications} \label{s4}
\begin{figure*}[tbh]
	\centering
	\includegraphics[scale=0.57]{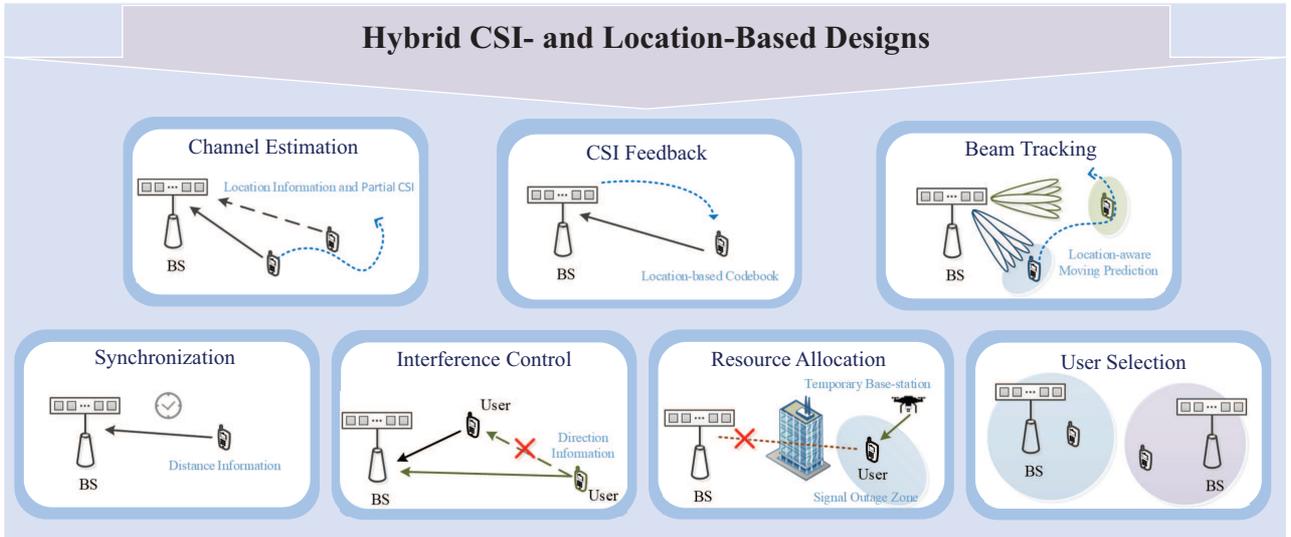}
	\caption{Location-assisted millimeter-wave communications} \label{fig:assist}
\end{figure*}
The electromagnetic properties in the mmWave frequency bands determine the high directionality of mmWave communications.
Therefore, the location information (which includes velocity) is connected with all aspects of mmWave communications, e.g., free space path-loss, Doppler shift, channel quality, beam directions, blockages, and interference levels.  
Traditional mmWave communications are operating totally based on the estimated CSI, thereby requiring highly frequent beam training and channel estimation processes to overcome the large path-loss and high blockage probability experience by mmWave signals, especially in high-mobility scenarios.
Motivated by the highly-accurate location information obtained by communication systems, the traditional CSI-based communication solutions can be turned into hybrid CSI-based and location-based solutions.
The location information can be harnessed to speed up and enhance beam training and tracking processes, which improves the CSI accuracy and reduces communication overhead.
Many aspects of mmWave communications can be enhanced with location assistance, as shown in Fig.\ref{fig:assist}.

\textbf{Channel Estimation. }
Although small scale-fading decorrelates over very short distances, the distance-dependent path-loss and the obstacle-caused shadowing correspond to the location of transceivers and obstacles.
This phenomenon implies that coarse CSI can be predicted from the side information from the location and environment, and this prediction can be complemented with instantaneous small-scale information, thereby reducing communication overheads.
The predicted CSI can be considered as a prior distribution
on the channel quality for possible uses.
The challenge is the requirement of flexible CSI predictive
engines, which should be able to learn and adapt in different radio propagation environments.

\textbf{CSI Feedback. }
Given that a coarse CSI can be predicted from the side information from the location and environment, the CSI feedback mechanisms can also be transformed from the state-of-the-art mechanisms
to the location-aware and environment-aware adaptive feedback mechanisms to reduce the feedback delay.
Feedback location information with some instantaneous channel information supplement can substantially reduce feedback overhead without compromising the data rate.

\textbf{Beam Tracking. }
Supporting ultra-fast and high-rate data exchanges
among moving users and between users and infrastructure can hardly be accomplished in traditional mmWave communications due to the unacceptable overhead.
By contrast, by using spatial movement coherence in combination with location prediction, the predicted coarse beam directions can be obtained to greatly narrow down the beam search area.
In particular, the location information can significantly speed up the beam tracking process with a low overhead.

\textbf{Synchronization. }
Synchronization
can be aided through a priori location information, which determines
the potential window to exploit the synchronization signals
from different base stations.

\textbf{Interference Control. }
The location of the transceivers also provides useful information about the interference level to be experienced at a specific location, time, and frequency. Hence, the moving user can accordingly adjust the transmission power, frequency, or beam direction to reduce interference from other users.
This process is very important in mmWave communication systems with densified and heterogeneous wireless networks, in which the system performance is largely constrained by interference.

\textbf{Resource Allocation. }
The channel capacity and outage can be predicted by using the spatial movement coherence in combination with location and channel prediction.
Techniques for location-assisted resource allocation can reduce overheads and delays beyond traditional time scales due to their ability to predict channel quality.
The communication rate is adapted to not exceed the predicted capacity, or extra spectrum resources can be pre-allocated to support the unexpected surge of communication traffic demand.
Temporary base-stations can also be used to provide communication links in unfavorable signal outage zones.

\textbf{User Selection. }
Location information is also beneficial in reducing the overhead associated with user selection mechanisms.
The easiest way is allowing the base stations to make decisions based solely on the users' locations.
Future trends will consider the terrains and layout of blockages, enabling the users to avoid or escape from unfavorable signal dead zones.

Location-awareness bears great promise to the mmWave communication revolution in terms of reducing delay and feedback overhead, improving link reliability, and maintaining high-throughput communications.
To maximize the use of such side information of locations, efforts are still needed in designing adaptive combination mechanisms to complement the location information with instantaneous CSI, investigating the performance tradeoffs between mmWave localization and communication, constructing advanced protocols for information interaction between localization and communication layers, and forming a new frame structure to allocate time-and-frequency resources for the localization function.

\section{Future Trends}\label{s5}

Current mmWave communication systems are designed for wireless communications, not for localization applications. 
Hence, high-throughput communication and high-accuracy localization via the integrated mmWave communication and localization systems requires additional studies

\subsection{Hardware Impairments}
At the mmWave frequency bands, risks of increased phase noise and non-linearity of communication systems exist.
These conditions could affect the signal quality and result in channel estimation errors and hence influence the functionality.
Hardware impairments should be considered in communication and localization algorithm design.
The uniqueness of the hardware impairments of each device can also be used for identification, thereby turning waste into wealth.

\subsection{Cross Layers}
The communication layer or localization layer are currently studied separately. 
Novel waveform designs should be devised to enable convergent communication and localization in mmWave systems, which efficiently satisfy trade-offs between the communications and localization requirements and share resources in the time, frequency and space domains.
Dynamic medium-access control protocols and radio resource management algorithms will be needed to allocate the radio resources according to the needs of different communication and sensing services.
A high degree of information sharing and adaptive combining mechanisms should be realized among communication and localization layers.

\subsection{Cross Devices}
No single technology can currently meet the requirements of ubiquitous communications, high-resolution localization, and energy-efficiency of future cellular networks while operating at high frequency range, high mobility, multiple targets, and clutter communication scenarios.
Therefore, multi-domain information collected from mmWave networks, sub-6GHz networks, sensor networks, WLAN, satellites, UAVs, and radar systems should be integrated by the upgrade of data fusion technology.
Interference management and cancellation techniques will be redesigned in such a heterogeneous network, together with the large difference in transmission powers needed for reliable communication and localization.

\section{Conclusions}\label{s6}
In summary, we have reviewed the background and explored the motivation of the integrated communication and localization in mmWave communication systems.
We have explicitly stated that the signal and channel models are the basis of communication and provides the necessary parameters and information for localization.
We explained the enhancement of mmWave communication system in both communication and localization in terms of several key enabling technologies, such as ELAA, RIS, and AI, together with the forthcoming challenges.
Then, we overviewed the basic knowledge of localization and proposed two concrete, feasible, and promising research directions for localization in mmWave communication systems.
These directions are localization with extremely-large antenna array and model-based neural networks, on the basis of the new hardware architecture and algorithm system, respectively.
We discussed the location-assisted mmWave communications by summarizing the opportunities in all aspects of mmWave communications, including channel estimation, CSI feedback, beam tracking, synchronization, interference control, resource allocation, and user selection. 
Finally, we concluded the future trends of the integration of communication and localization in mmWave systems, which contain hardware impairments, system designs, and algorithm updates, for cross communication and localization layers and various devices in heterogeneous networks.

\bibliographystyle{IEEEtran}
\bibliography{bibsample}

\end{document}